# Big data analyses reveal patterns and drivers of the movements of southern elephant seals

Jorge P. Rodríguez[a], Juan Fernández-Gracia[b], Michele Thums[c], Mark A. Hindell[d], Ana M. M. Sequeira[e], Mark G. Meekan[c], Daniel P. Costa[f], Christophe Guinet[g], Robert G. Harcourt[h], Clive R. McMahon[i], Monica Muelbert[j], Carlos M. Duarte[k], Víctor M. Eguíluz[a] a: Instituto de Física Interdisciplinar y Sistemas Complejos IFISC (CSIC-UIB), E07122 Palma de Mallorca, Spain
b: Department of Epidemiology, Harvard T.H. Chan School of Public Health, Boston MA, USA
c: Australian Institute of Marine Science, Indian Ocean Marine Research Centre, University of Western Australia (M470), 35 Stirling Highway, Crawley, Western Australia 6009, Australia
d: School of Zoology, University of Tasmania, Private Bag 05, Hobart, Tasmania 7001, Australia
e: IOMRC and The UWA Oceans Institute, School of Animal Biology, University of Western Australia, M470, 35 Stirling Highway, Crawley, Western Australia 6009, Australia
f: Department of Ecology & Evolutionary Biology, University of California, Santa Cruz, California 95060, USA
g: Centre d'Études Biologiques de Chizé, UMR 7372 CNRS-Université de La Rochelle, 79360 Villiers-en-Bois, France
h: Department of Biological Sciences, Macquarie University, Sydney, New South Wales 2109, Australia
i: Sydney Institute of Marine Science, 19 Chowder Bay Road, Mosman, New South Wales 2088, Australia
j: Instituto de Oceanografia, Caixa Postal 474 - 96201-900, Rio Grande, RS Brasil
k: Red Sea Research Center (RSRC), King Abdullah University of Science and Technology (KAUST), Thuwal, 23955-6900, Saudi Arabia

**Abstract**
The growing number of large databases of animal tracking provides an opportunity for analyses of movement patterns at the scales of populations and even species. We used analytical approaches, developed to cope with "big data", that require no 'a priori' assumptions about the behaviour of the target agents, to analyse a pooled tracking dataset of 272 elephant seals (*Mirounga leonina*) in the Southern Ocean, that was comprised of >500,000 location estimates collected over more than a decade. Our analyses showed that the displacements of these seals were described by a truncated power law distribution across several spatial and temporal scales, with a clear signature of directed movement. This pattern was evident when analysing the aggregated tracks despite a wide diversity of individual trajectories. We also identified marine provinces that described the migratory and foraging habitats of these seals. Our analysis provides evidence for the presence of intrinsic drivers of movement, such as memory, that cannot be detected using common models of movement behaviour. These results highlight the potential for "big data" techniques to provide new insights into movement behaviour when applied to large datasets of animal tracking.

**Introduction**

Movement is a fundamental aspect of animal behaviour[1]. The need to search for food, mates and shelter shapes many aspects of animal ecology and is central to developing conservation and management strategies for any species[2,3]. Studies of animal movement were catalysed by the introduction of satellite-linked tags and the Argos satellite system in the late 1970's[4,5], which for the first time allowed animals to be tracked in a near-real time across habitats such as the forests, skies and open oceans that had previously been largely inaccessible to researchers.

Observations describing horizontal displacements have been the most common product of satellite-linked tags. Analysis of these tracks can reveal the processes that underlie the movement strategies of the target species[6] and have mostly focused on the role of prey distribution in determining movement patterns[7-9]. However, movement patterns are unlikely to be solely a response to the spatial and temporal distribution of food[10,11]. Animals have the capacity to learn and react to important aspects of their environment for many reasons, such as reproduction and anti-predator behaviour[12], or even fear[13,14]. Some movement behaviours may even be genetically programmed[15,16]. Examination of these ideas has been limited in the past by the small sample sizes of most tracking studies due to the expense of satellite tags. Such low replication led to mostly individual-based analysis rendering any intrinsic (learning, genetic) component of movement behaviour difficult to detect. In recent years, satellite tagging has become more widespread not only in research but also as a cost-effective monitoring tool, so the limitation of small sample size can now be overcome through pooling data across multiple studies[2], creating large datasets of movement. Using new powerful computational resources, these large datasets can now be subject to powerful numerical and analytical approaches capable of identifying collective movement patterns, such as those previously used in studies of human mobility[17-19]. Such analytical tools offer the opportunity to examine how animals utilise space both at the level of individuals and that of populations and species, thereby identifying the roles of intrinsic drivers of movement patterns.

Here, we use analytical tools originally developed for the analysis of 'big data' produced by studies of human mobility to explore movement patterns of southern elephant seals. These seals are recognized as a keystone predator within cool-temperate and Antarctic food chains, and an understanding of the drivers of their movement patterns is essential, given they are likely to be strongly affected by anthropogenic threats such as global warming[20]. Southern elephant seals are an ideal candidate with which to explore variability in movement patterns and space occupancy in a data-centric approach, as earlier studies have compiled large datasets composed of hundreds of individual tracks. Such large sample sizes reflect aspects of the life history of elephant seals that make them amenable to tracking studies, since they are large, long-lived animals capable of carrying tracking instruments with large storage and processing capacity. They are also long-distance (100s-1000s km) migrants across open oceans, showing fidelity to colony locations on land for breeding and moulting, offering the opportunity to deploy and retrieve satellite-tracking devices so that archives of high-frequency sampling can be recovered from tags. The aim of our study was to search for unifying patterns in the space use of elephant seals, from the scale of individuals to the entire species. In so doing, we sought to identify and quantify both extrinsic and intrinsic drivers of movement patterns of these animals.

**Results**
We analysed a dataset of 550,537 individual locations obtained from Argos platform transmiting terminals (PTT) deployed on 272 southern elephant seals (SES) between 2004 and 2013 at seven different locations in the Southern Ocean (Fig. 1, see Supplementary Fig. S1 online). The probability density function (pdf) of their aggregated displacements displayed a universal shape across several spatial and temporal scales characterized by a power-law scaling regime. After rescaling, dividing each displacement by the average displacement $D=\frac{d}{\langle d \rangle}$, the pdf of displacements for a given time window T, $p(d;T)$, led to $P(D)$:

$$p(d;T) = \langle d(T) \rangle^{-1} P(D)$$

(1)

where $P(D)$ had a universal shape, described by a power-law with a sharp cut-off (Fig. 2A, see previous approaches for animal movement analysis in Supplementary Fig. S2 online[21,22]). For displacements shorter than the average displacement, *i.e.*, where $D<1$, the behaviour was described by a power-law decay $P(D) \sim D^{-\gamma}$, with the exponent $\gamma= 0.60$ that characterized the movement. For larger displacements, $D>1$, the pdf decayed abruptly. The scale-free behaviour observed indicated that the elephant seals used the same movement strategy across multiple spatial scales up to a characteristic distance corresponding to the maximum travel speed of the species. Both the average displacement and the mean square displacement scaled as a power of time, $\langle d \rangle \sim T^a$ and $\langle d^2 \rangle \sim T^b$ respectively, with exponents $a=b/2=0.83$ (see Supplementary Fig. S3 online), above the value known to correspond to Brownian motion ($a^{dif}=1/2$). Such scaling exponents are characteristic of directed movement.

Further in-depth inspection of movement patterns revealed that the scale-free behaviour of elephant seals at micro-scales (*i.e.* statistical features of the displacements) translated into a complex pattern at macro-scales (spatial densities). Our site occupancy, that is the number of locations in a grid cell per unit area, analysis based on the aggregated dataset showed that a large fraction of the area used by the seals was characterized by low occupancy, while a small fraction of grid cells had high occupancy (Fig. 1B). The low occupancy areas corresponded to long displacements conducted at relatively high speeds, while high occupancy areas were characterised by short displacements at low but highly variable speeds (Fig. 2B). The two pdf's of displacements based on occupancy revealed that only high-occupancy areas retained a distribution characterised by a power-law with an exponent of 1.17. Movements in these high-occupancy areas were thus responsible for the scale-free nature of the overall distribution of displacements. The occupation densities $\rho_{ev}$ were broadly distributed, with a pdf that displayed a power-law decay with exponent 1.88, making 80% of the observations occur in the 23.7% of grid cells that received the most visits (see Supplementary Fig. S4 online).

The general movement laws described above emerged from the aggregation of behaviours of individual southern elephant seals. Focusing on the individual tracks, we analysed the spatial spread of observed trajectories, the impact of displacement correlations in spatial analysis, and the fidelity to particular grid cells. We found that the gyration radius, an indicator of the spatial dispersion of individual trajectories, ranged from <10 to 2000 km, highlighting the enormous variation in spatial dispersion among individuals (Fig. 3).

Previously, we showed that SES movement has two distinct main modes, described in high and low occupied regions (Fig. 2B), with clear evidences for directed movement (see Supplementary Fig. S3 online). These displacement features suggest the presence of correlations in the sequences of both displacements and turning angles. In order to test this, we compared observed individual trajectories with a reshuffling of them to break these correlations (see Methods and Supplementary Fig. S5 online), finding that the patterns of space use for actual trajectories included a higher number of visited grid cells than for the reshuffled trajectories (Fig. 4A inset). This means that the observed trajectories lead to a more extended exploration of space than would be the case in trajectories without correlations. We measured the fidelity of an individual to particular grid cells with the entropy of its pattern of spatial visitation $S$, so that frequent visits to the same area led to low values of entropy, while uniform random visitation led to large values[23,24]. The distribution of the entropy for the trajectories of individual SES revealed high variability with most trajectories showing large entropy ( > 0.6), which indicates a relatively uniform probability of visiting each grid cell of the trajectory; hence, most SES did not make repeated visits to particular grid cells (see Supplementary Fig. S6 online). From the entropy $S_i$ and the number of visited cells $M_i$ of an individual $i$, we calculated its limit of predictability $\Pi_i^{MAX}$ (see Methods). The distribution of limits of predictability showed a smooth decay from the maximum around 0.2 to 0.8 (Fig. 4A), with 60% of the individual trajectories represented in the limit of predictability range between 0.2-0.4. However, our analysis also revealed some limit of predictability values close to 1, indicating that the corresponding trajectories were indeed highly predictable. In fact, longer trajectories (measured in terms of the number of visited grid cells) led to a uniform probability to visit each grid cell and thus to a low limit of predictability, whereas short trajectories led to fidelity to a few areas of the grid and then had a high limit of predictability (Figs. 4B-D). This analysis revealed high levels of heterogeneity among trajectories, highlighting again the range of individual variation in the movements of seals. Although most limits of predictability were low, we found that the limit of predictability for the reshuffled trajectories averaged at 0.18 ± 0.07 (standard deviation), having smaller values than in the observed trajectories for most individuals.

Collective movements revealed marine provinces[25,26], geographical areas used consistently by several elephant seals from different sub-populations. In this analysis, we applied community detection methods to the transition probability matrix obtained from the trajectories, in which entries represented the flux between two cells, that is, the fraction of trips coming from one cell that ended up in another (see Methods). The community detection software identified a hierarchy of provinces (aggregated grid cells) from the most (level 0) to the least inclusive (numbered in successively from 0). For transitions after a day ($T$ =1), we found two provinces at level 0, and six at level 1 (Fig. 5A). The provinces also characterized the mobility range of individual seals given that 80% of the seals spent more than 80% of their time in a single province (Fig. 5B). Seals from Elephant and Livingston Islands had an overlapping province (light blue), as did those from Macquarie, Campbell and Livingstone Islands (dark blue), while the seals tagged at Kerguelen Island, Casey and Davis Stations shared four provinces (red) (Fig. 5C).

**Discussion**

The integrated analyses used here characterised both individual and collective movement behaviours of elephant seals, a key top-order predator in the Southern Ocean. We found that scale-free signatures of movement patterns emerged from these analyses were indicative of search strategies likely related to prior knowledge of the location of foraging grounds, thus providing evidence that memory is likely to be an intrinsic driver of the movement.

The resulting pdf of the aggregated displacements was described by a power law with an exponent smaller than 1. Power-law distributions are characterized by scale-freeness, leading to scale invariance: elephant seals used the same strategy to search their environment not only across many spatial but also across temporal scales, as evidenced from the collapse of the distribution function for different temporal resolution used to measure displacements. In the controversial[27] Lévy foraging (LF) hypothesis, scaling exponents close to 1 are argued to occur in situation of sparsely distributed resources[7]. Alternatively, the probability of return to a breeding site may control the scaling exponent of probability function distributions[28], with measures of entropy showing a maximum at a scaling exponent that shifts from 2 to 1 as the probability of return increases. This situation is likely to occur in elephant seals, given that the majority of tracks we analysed were return journeys between breeding colonies and foraging grounds. When areas of high occupancy were analysed separately, seals had trajectories characterised by short displacements at low speed, which were likely to be representative of foraging, with a scaling exponent bigger than 1. Such behaviour is also consistent with area-restricted search[29]. In contrast, trajectories in grid cells that were infrequently occupied included longer displacements that occurred at speeds twice the average rate. Thus, our results support the idea that these wide-ranging predators combine deterministic movement over very large (100-1000s km) spatial scales[11,30] with more probabilistic movement over smaller (10-100 km) spatial scales[31]. Our evidence for such combinations of behaviour might also account for the vertical movements of elephant seals[32] since it is known that these deep diving animals (average dives of 300 m at night and 600 m during the day) also target specific foraging depths[33].

The analytical techniques we applied offer a number of advantages compared to alternative approaches[21,22,32]. We were not required to define turning points in order to resolve steps in the data, removing *a priori* assumptions about the movement of the individuals. This technique also allowed the assessment of differences in movement behaviours at different time scales and importantly, enabled the description of an entire movement strategy, rather than just the movement assumed to correspond to foraging. The latter has been a principal goal of many studies of animal movement and much of this work has focused on the Lévy foraging hypothesis[32,34,35]. This hypothesis contends that Lévy walks are optimal search strategies for animals over a very broad range of foraging conditions, most typically where food is scarce and unpredictable[36]. This has led to a focus on resource abundance and distribution (and the environmental factors that determine this phenomenon) as drivers of the movement patterns of marine predators[7]. Evidence for this hypothesis is, however, equivocal; reviews of the literature have argued that Lévy walks are ubiquitous in many marine species[32], whereas others have argued that the hypothesis does not adequately describe movement patterns of some animals. The Levy foraging hypothesis has also been challenged from several perspectives, for reasons including the unrealistic nature of underlying model and the lack of optimality at relevant spatial scales[37], among other criticisms[27]. As noted above, our analysis departs from the Levy foraging hypothesis, since we do not identify 'turning points' indicative of behaviour or assume an underlying Lévy-like behaviour. However, the universal shape of the pdf of displacements found here supports the hypothesis that it is a signature of the movement pattern. Indeed, the robustness of the displacement distribution across temporal scales with a scaling exponent (less than 1) was outside the stable regime according to the central limit theorem. This result was a signature of non-Markovianity and suggests that memory was a driving force of movement patterns[38].

Occupancy and entropy of spatial patterns of visitation were used to characterise the fidelity of elephant seals to particular grid cells during a track[23,24]. The scale-free nature of the distribution of occupancy, the hierarchy of occupation, ranging from grid cells that were occupied frequently and for a long duration to others that were occupied briefly and rarely, were a signature of the complex movement patterns we observed. The entropy of the individual trajectories had a broad distribution, revealing that the seals tendency to occupy areas varied widely among individuals. The distribution of individual entropies peaked at a large value of around 0.9, indicating that most individuals visited different parts of the grid with a probability distribution close to uniform (*i.e.*, random; S=1), so that the limit of predictability was low (mostly between 0.2 and 0.4; Fig. 4), particularly when compared to patterns of human movement ($\Pi^{MAX} = 0.9$)[39]. However, the limit of predictability of reshuffled movement data of elephant seals was even lower (0.18 ± 0.07 SD; Fig. 4), pointing out that visitation patterns deviated from random, a result consistent with previous observations[10,11,31]. The idiosyncratic and temporally variable nature of the movement patterns of animals can make it difficult to characterise movement at the scale of populations or species. When trajectories were analysed on an individual basis, the distribution of the gyration radii revealed a large range of characteristic spatial scales among individuals (from <10 km up to 2000 km). Low gyration radii were recorded for individuals staying close to tag deployment locations, while the highest radii were recorded for individuals undergoing long distance migrations, independent of tag deployment location (Fig. 3). While individual seals show idiosyncratic behaviours making it challenging to formulate generic descriptions of movement at the species level, the combination of the results we obtained for occupancy, entropy and gyration radius suggests that a wide range of movement patterns is related to the presence and distance of foraging grounds relative to the location of the colony sites, where the animals aggregate to breed, moult or rest.

Previous approaches for the spatial representations of animal use describe individual animal tracks, so they are difficult to use in population context, which requires some type of statistical aggregation of individual tracks. Our application of community detection techniques to the transition probability matrix given by aggregate mobility patterns provides an automatic, widely applicable and computationally-easy means of dividing the movement space into relevant sections or provinces dictated by the mobility patterns of the study species, and can even be applied to flows with memory[40]. Our community detection analysis identified geographical borders between areas that showed different use patterns by the seals encompassing the routes that southern elephant seals used to travel from their colonies to their foraging grounds and back. These methods provided the most balanced distribution of province size and connectivity at a time window of *T*=1 day (Fig. 5). The fact that low occupancy, transiting areas were included in the provinces associated with each colony highlights the general utility of this method for identifying migration corridors between the home colonies and foraging grounds. This is important because such corridors are typically disregarded by traditional methods for estimating space use, such as time in area approaches or kernel densities. The provinces identified by our approach agree well with what is already known for this species. For example, the analysis captures the migratory regions for each population. However, as the seals are wide-ranging, there is considerable joint usage of geographic regions among the populations, and our new approach is able to easily account for this, clearly identifying provinces that are universally migration corridors, or foraging regions. Previous studies have described distinct foraging strategies among populations. For example, at Kerguelen Island, seals either use sub-Antarctic foraging grounds or high-Antarctic foraging grounds[41], and there is a similar division at Macquarie Island. These regions appear to yield contrasting energy gains, due to differing habitat quality, and this seems to play an important role in population trajectories. Our big data approach provides a quantitative basis for identifying these differing provinces which will enable rigorous development and testing of hypotheses regarding foraging decisions.

Through the application of big data techniques developed for studies of human movement to the tracking of elephant seals, we found evidence that at large spatial scales, southern elephant seals do not behave in ways consistent with the assumptions of analytical approaches commonly used in animal ecology. Such approaches tend to be based on the idea that animal movement decisions are made purely on some current assessment of environmental and/or resource conditions, principally the density of prey. We argue that such conceptual framework is incomplete, because movements are, at least in part, likely to be associated with some prior experience of the location of prey (*i.e.* memory), and with activities other than foraging. Moreover, our approach also highlights the ability of the analyses to quantitatively depict the idiosyncratic behaviour of individual elephant seals, by describing plasticity of movements in relation to the varying distances to foraging areas and position of the colony site. Finally, the marine province analysis revealed that elephant seals partitioned space into consistent units of use, or provinces, which encompassed corridors for migration and foraging locations. Such information is fundamental for ecological spatial planning and management tasks, as we were able to identify provinces including not only moulting or breeding locations, but also corridors between them, and we detected highly diverse areas in regions including several marine provinces. Thus, our analytical approaches not only provide a new framework for describing and classifying the use of space by marine animals, it also offers insights into the likely drivers of these patterns of movement.

**Methods**
**Data.** The dataset of locations was obtained from Argos platform transmiting terminals (PTT) deployed on 272 individuals at seven locations in the Southern Ocean (15 southern elephant seals tagged at Macquarie Island, 11 at Campbell Island, 97 at Kerguelen Island, 29 at Elephant Island, 52 at Livingston Island, 24 at Casey Station and 44 at Davis Station (Antarctica)) between 2004 and 2014 (Fig. 1A, see Supplementary Fig. S1 online). All deployments were made at the end of the seal annual breeding haul-out (prior to the post-breeding migration) or at the end of the annual moult haul-out (prior to the post-moult migration). Macquarie and Kerguelen Islands are breeding and moulting locations for the seals of both sexes, whereas Casey and Davis Stations are only moulting (and resting) locations for predominantly non-breeding seals that are also predominantly male. The dataset also included adult and juvenile females and juvenile and sub-adult males. For tagging, seals were chemically sedated[42], weighed and measured[43] and a Sea Mammal Research Unit (University of St Andrews) satellite relay data logger (SRDL) was glued to the back or head of the seal. The combined weight of the tags and glue was approximately 0.5 kg, *i.e.* 0.15% and 0.10% of the mean departure weight of adult female southern elephant seals (338 ± 65 kg) and sub-adult males (469 ± 202 kg), respectively. We are confident that the instruments did not affect at-sea-behaviour given that the smallest instrumented seal weighed 169 kg (<0.3% of the seals' weight). Previous studies have demonstrated that seals carrying twice this load (instruments up to 0.6% of their mass) were unaffected in either the short-term (growth rates) or the long-term (survival) by carrying these instruments[44].

Seal movements at-sea were determined using the ARGOS satellite tracking system, which uses the Doppler shift in transmitted frequencies to estimate animal position. Positions are subsequently classified into one of seven location classes (LC 3, 2, 1, 0, A, B, and Z) that have a $68^{th}$ percentile spatial error ranging from 0.5 km (LC 3) to 36 km (LC B)[45]. Location and quality of the location estimate were provided for each uplink. The SRDLs remained on the seals until they either fell off or were shed with the hair during the next annual moult. State-space models[46] were used to minimize positional errors and to estimate location points along movement paths at two-hour time steps. All southern elephant seal data used in these analyses was collected by a large team of investigators and has previously been published in some form[47-51].

All procedures were approved by the respective ethics committees and licensing bodies including, the Australian Antarctic Animal Ethics Committee (ASAC 2265, AAS 2794, AAS 4329), the Tasmanian Parks and Wildlife Service, the University of California, Santa Cruz and the Programa Antártico Brasileiro, and were carried out in accordance with current guidelines and regulations.

**Probability density function (pdf) of displacements.** Displacements were calculated by measuring the geographic distance (great circle distance) $d_{i,t}(T)$ between two positions of the same individual trajectory $i$ at time $t$, that were separated by a time window $T$, i.e., we measured the distance between locations at $t$ and $t+T$. We then aggregated all the displacements from all the individual tracks for each time window considered to obtain the pdf of displacements for each time window. To compare the functional shape of the pdf at different time windows, we rescaled the displacements, dividing each displacement by the average displacement $\langle d(T) \rangle$, $D = \frac{d(T)}{\langle d(T) \rangle}$. The average displacement was calculated for the corresponding time window $T$: $\langle d(T) \rangle = \frac{1}{C} \sum_{i,t} d_{i,t}(T)$, where $C$ is the total number of displacements of duration $T$, $i$ represents the runs for all the considered individuals and $t$ is the time. The root-mean-square-displacement, for a given time $T$, was calculated with the square root of the average of the square displacements occurring after a time window $T$, $d_{RMS} = \sqrt{\langle d^2(T) \rangle}$. We checked that different number of locations per individual did not influence the aggregate statistics, with an alternative analysis in which every individual had the same contribution to the aggregated pdf, irrespective of its number of locations in the dataset, obtaining results with no significant differences.

**Fitting power-law distributions.** We used the maximum likelihood estimation (MLE) method[52,53] for fitting the exponents of the power-law distributions. For Fig. 2A, we chose a truncated Pareto distribution, $f(D) = \frac{1-\gamma}{D_{max}^{1-\gamma} - D_{min}^{1-\gamma}} D^{-\gamma}$, between the minimum displacement in our data ($D_{min}$) and $D_{max}=3$, and finding the value of $\gamma$ which maximized the log-likelihood function with numerical methods. For Supplementary Fig. S4, we fitted a truncated Pareto distribution between $\rho_{min}=0.3$ km$^{-2}$ and $\rho_{max}=100$ km$^{-2}$.

**Discretization of space.** We discretised the space in grid cells of resolution 0.25° × 0.25°. Each grid cell $i$ was then characterized by its occupancy $\rho_i = \frac{\sum_j n_{i,j}}{A_i}$, where $n_{i,j}$ is the number of locations of individual $j$ that fell in grid cell $i$, $j$ represents for all individuals, and the area of that grid cell is $A_i$. The resolution of 0.25° was chosen after examining the pdf of grid cell occupancy density for different resolutions, where we found a delta function for low resolutions (all the events located in a single grid cell) and also for very high resolutions (all the visited grid cells are visited once). We considered that the suitable resolution for this analysis was the one far from the limits leading to a pdf of occupancies displaying delta functions, according to our data (see Supplementary Fig. S4 online). We then ranked the grid cells according to the occupancy and considered the first third of the ranking to be highly occupied grid cells and all the others as low occupancy. We chose the first third of ranked grid-cells to compensate for the larger number of observations occurring at a few highly occupied locations (a consequence of the scale-free distribution of occupancy).

**Gyration radius.** Using each individual track, we calculated the centre of mass of each trajectory by converting the observed locations from cylindrical to Cartesian coordinates with the origin at the Earth centre, and then calculating the average position of the individuals. We then calculated the gyration radius[17] (*i.e.*, the dispersion of the observed positions), measured as the standard deviation of the distances from every location to the centre of mass (Fig. 3): $r_G = \frac{1}{N}\sqrt{\sum_{i=1,N}[\text{dist}(\vec{x}(t_i), \vec{x}_{CM})]^2}$, where $\vec{x}(t_i)$ is the position on the Earth surface of the seal at time $t_i$, $\vec{x}_{CM}$ is the centre of mass of the trajectory calculated by averaging the position in a three dimensional sphere, and projecting the average in the sphere surface; and $\text{dist}(\vec{y}, \vec{z})$ is the distance between $\vec{y}$ and $\vec{z}$ along the greatest circle connecting both points on the sphere.

**Reshuffling of trajectories**. From the observed trajectories we reshuffled the sequence of displacements in such way that they kept the same pdf of displacements but without correlations. For each individual, between each position and the next, we calculated the projected distances of the movement on directions N-S (y axis) and W-E (x axis), obtaining the change in latitude $\Delta y$ and the change in longitude $\Delta x$, *i.e.*, a vector $(\Delta x, \Delta y)$ for each time with a recorded location. Keeping the same origin point, we obtained a sequence of vectors that we then randomized. The end point is defined by the origin point plus the sum of all the vectors, and therefore, the order of the sequence did not change the endpoint (see Supplementary Fig. S5 online).

**Entropy of trajectories.** The probability that an individual $j$ visited cell $i$, that is, the fraction of data points from that seal's trajectory located in that cell, $p_j(i)$, was used to compute the entropy of that trajectory defined as $S_j = -\sum_i p_j(i)\log(p_j(i))$, where the sum runs over all visited cells. Given that individual $j$ visited $M_j$ areas, we normalized the entropy of its trajectory by the entropy that corresponds to a uniform visitation probability $p_{\text{unif},j} = \frac{1}{M_j}, S_{\text{unif},j} = -\sum_{i=1}^{M_j} -\frac{1}{M_j}\log(M_j) = \log(M_j)$.
This normalization allows for direct comparison of the entropies of trajectories with different numbers of visited areas and informs about the complexity of the visitation pattern ranging between 0 (one visited cell) and 1 (uniform, every cell is visited with the same probability).

**Predictability.** We calculated the limit of predictability $\Pi^{MAX}$, a measure of the theoretical maximum probability to predict the location of a trajectory[39], based on estimated entropy ($S$) and the number of visited cells($M$). $\Pi^{MAX}$ was, therefore, obtained solving the following implicit equation:
$$S = H(\Pi^{MAX}) - (1 - \Pi^{MAX})\log(M-1) ,$$
(2)
where $H(x) = -x\log(x) - (1-x)\log(1-x)$.

**Provinces and community detection algorithms.** We identified the spatial areas of use for elephant seals, which we refer to as 'provinces', based on the transition probability matrix. Each element of this matrix, $\Omega_{ij}$, measured the probability of going from the grid cell $i$ to $j$ after a specific time window $T$. The element $\Omega_{ij}$ was then the number of those dyadic interactions representing the visitation patterns from $i$ to $j$ divided by the total number of visitation patterns from $i$ to any other cell within the prescribed time window $T$. Therefore, our transition probability matrix described a weighted directed network of the grid cells accounting for situations when the trajectories remained in the same grid cell or returned to it within the time window $T$ (self-loops). We specified $T$=1 day to identify marine provinces describing the movement of elephant seals with the community detection algorithm *Infomap*[54]. *Infomap* makes use of random walkers to explore a network and determines the existence of 'communities' by minimizing the information needed to describe a walker's trajectory. In that way, regions where the random walker remained for considerable amounts of time were identified as 'communities', without the need for prior information. This method works hierarchically, *i.e.* by finding different levels of communities that have different inclusiveness. We started our analyses from the most inclusive level (level 0), and the partitions were increasingly smaller as we progressed in the hierarchy, mimicking the shape of a phylogenetic tree. We show results for the two most inclusive levels of hierarchy only: 0 and 1 (Fig. 5).

**Acknowledgments**
Data from Macquarie Island, Davis and Casey Stations was sourced from the Integrated Marine Observing System (IMOS) - IMOS is a national collaborative research infrastructure, supported by Australian Government. The Australian Antarctic Division provided logistical support for those deployments. A.M.M.S. was supported by an IOMRC (UWA/AIMS/CSIRO) collaborative Postdoctoral Fellowship (Australia) and by ARC grant DE170100841; J.P.R. acknowledges support by the FPU program of MECD (Spain); J.F.G. is supported by NIH grant U54GM088558-06 (Lipsitch); M.M. acknowledges support from CNPq; V.M.E. acknowledges support from SPASIMM (FIS2016-80067-P (AEI/FEDER, UE)). Research reported in this publication was supported by research funding from King Abdullah University of Science and Technology (KAUST).

**Author contributions**
J.F.G, M.T., A.M.M.S, M.G.M., C.M.D. and V.M.E. conceived and designed the research; J.F.G, J.P.R., M.T., M.A.H., A.M.M.S, M.G.M., C.M.D. and V.M.E. analysed the data and discussed the results. J.P.R., J.F.G., M.T., M.A.H., A.M.M.S., M.G.M., D.P.C., C.G., R.G.H., C.R.M., M.M., C.M.D. and V.M.E. contributed to write the manuscript.

**Competing financial interests**
The authors declare no competing financial interests.

**Figures**

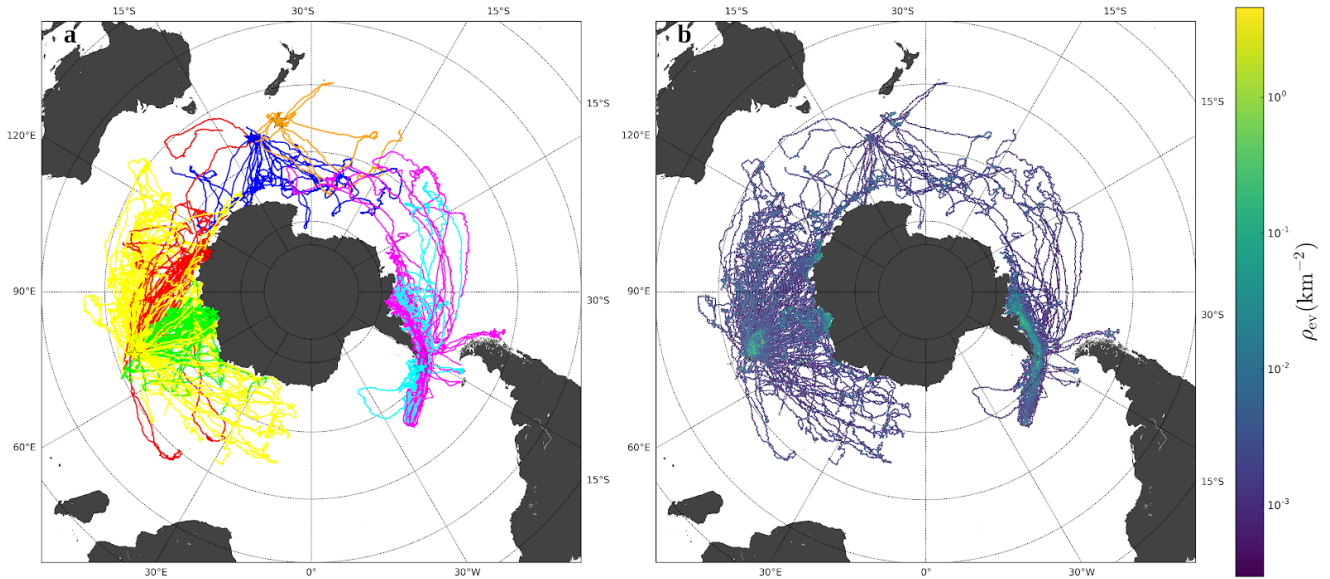

**Fig. 1. Description of southern elephant seals trajectories. a**, Map of all SES's trajectories. Land is shown in dark grey. Different colours correspond to elephant seals tagged at seven different deployment locations, which are represented with star symbols: Kerguelen (yellow), Macquarie (blue), Campbell (orange), Livingston (magenta) and Elephant (cyan) Islands, Casey (red) and Davis (green) Stations. **b**, Occupancy map of the trajectories. The colour scale is logarithmic, from violet (low occupancy) to yellow (high occupancy). Maps generated with Matplotlib Basemap Toolkit[55].

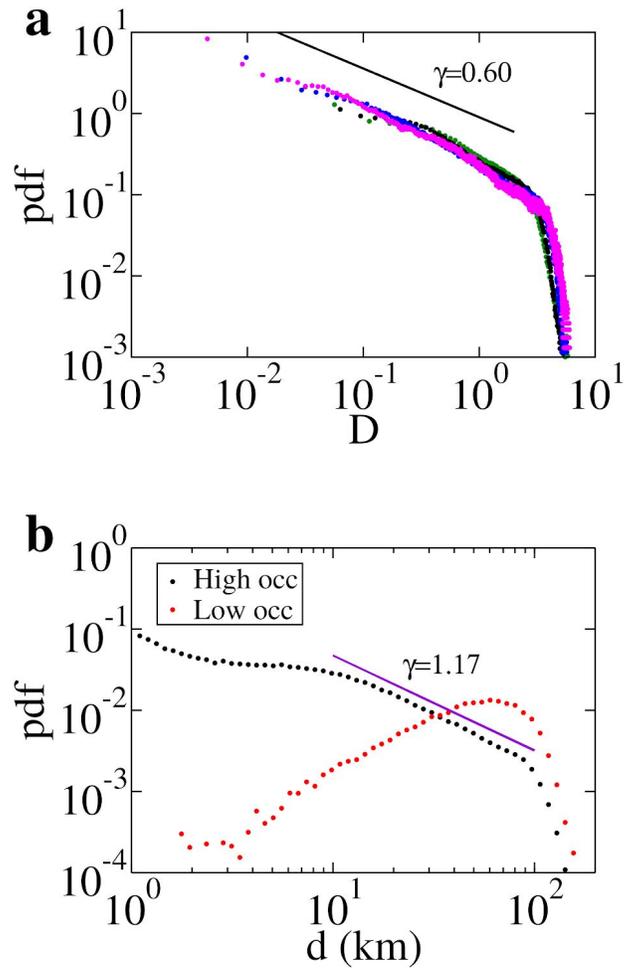

**Fig. 2. Analysis of displacements. a**, Probability density function (pdf) of normalized displacements *D* for 0.5 day (green), 1 day (black), 4 days (blue), and 10 days (magenta). The pdf's collapse into a universal function: for displacements below the average (*D*<1), the probability decays as a power-law with exponent $\gamma = 0.60$, while for larger displacements, the pdf decays abruptly; **b**, Distribution of displacements *d* for origins located at both high (black) and low (red) occupancy grid cells with *T*=1 day.

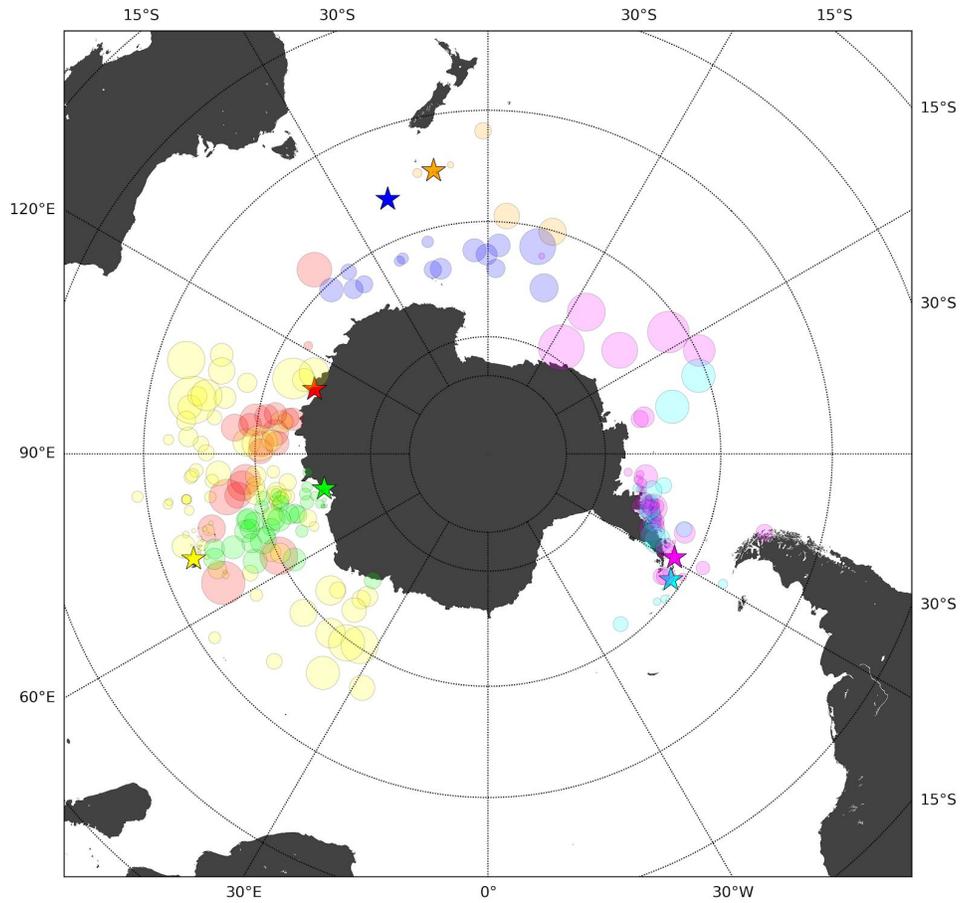

**Fig. 3. Spatial extent of the trajectories.** Map of the position of the centre of mass of each trajectory. Symbol size is proportional to the gyration radius $r_G$; colours indicate different deployment locations (associated with different populations) which are represented with star symbols: Kerguelen (yellow), Macquarie (blue), Campbell (orange), Livingston (magenta) and Elephant (cyan) Islands, Casey (red) and Davis (green) Stations. Map generated with Matplotlib Basemap Toolkit[55].

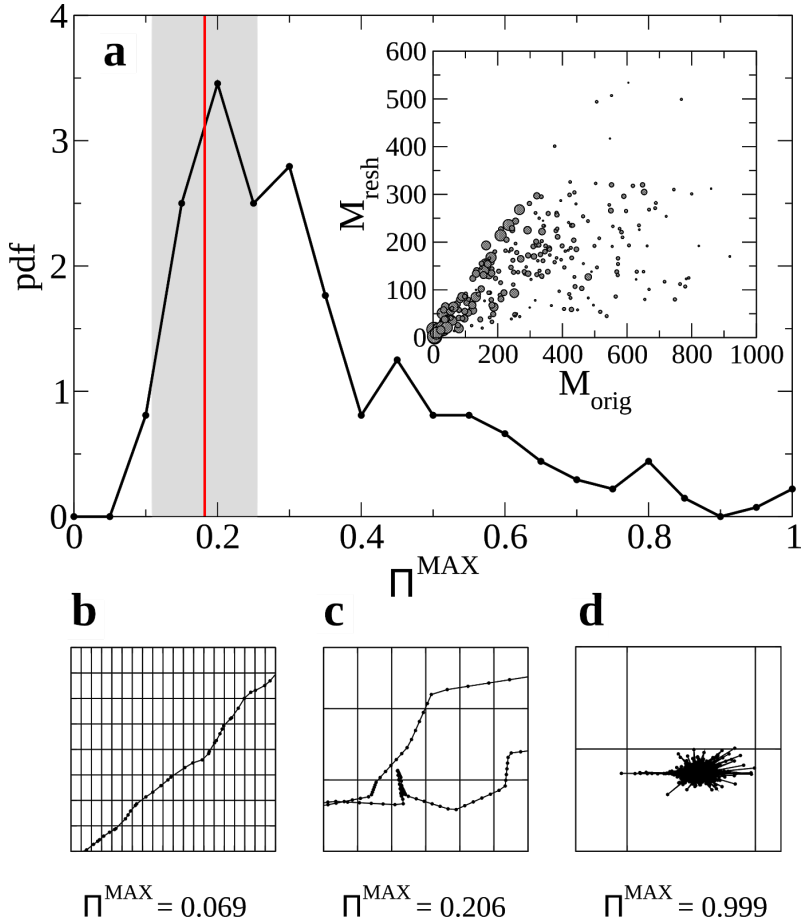

**Fig. 4. Analysis of the use of space. a**, Distribution of the limit of predictability, $\Pi^{MAX}$. The red line indicates the average of the maximum predictability distribution for reshuffled trajectories, and the shaded area represents the range of the limit of predictability obtained for the reshuffled trajectories within a standard deviation from the average. Inset: Number of visited grid cells, for each individual, for the reshuffled ($M_{resh}$) and the observed trajectories ($M_{orig}$). Symbol size represents the limit of predictability of the individual trajectory. The plots **b-d** depict typical trajectories whose values for the limits of predictability are 0.069, 0.206, and 0.999, respectively.

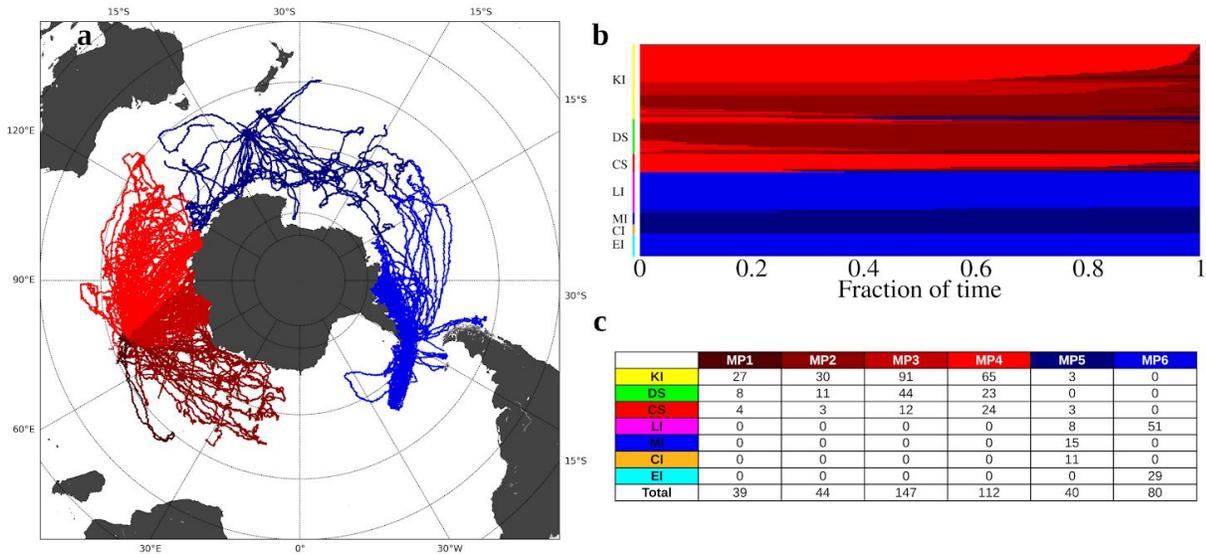

**Fig. 5. Marine provinces of southern elephant seals in the Southern Ocean.** Marine provinces were obtained based on the movement of elephant seals for the transition probability matrix $\Omega$ for time window $T$=1 day. **a**, Map of the marine provinces. Red and blue colours indicate the communities at level 0, while the darkness of the colours separates communities at level 1. Map generated with Matplotlib Basemap Toolkit[55]; **b**, Fraction of time spent at a given province per individual. Each row corresponds to a single animal, which was tagged at the locations indicated by the colour bar: Kerguelen Island (KI, yellow), Davis Station (DS, green), Casey Station (CS, red), Livingston (LI, magenta), Macquarie (MI, blue), Campbell (CI, orange) and Elephant (EI, cyan) Islands. The rows were divided into the colours of the provinces that each seal uses, and their lengths are proportional to the time spent in them; **c**, Number of elephant seals visiting each marine province (MP) for every deployment location; marine provinces were ordered according to **a**.

**Big data analyses reveal patterns and drivers of the movements of southern elephant seals**

Jorge P. Rodríguez, Juan Fernández-Gracia, Michele Thums, Mark A. Hindell, Ana M. M. Sequeira, Mark G. Meekan, Daniel P. Costa, Christophe Guinet, Robert G. Harcourt, Clive R. McMahon, Monica Muelbert, Carlos M. Duarte, Víctor M. Eguíluz

**Supplementary Information (SI)**

**SI Figure legends**

**Figure S1.** Observation time series for each individual. Different colours correspond to elephant seals tagged at seven different deployment locations: Macquarie Island (blue), Campbell Island (orange), Kerguelen Island (yellow), Casey station (red), Davis station (green), Livingston Island (magenta), and Elephant Island (cyan).

**Figure S2.** Step length analysis of the trajectories. **a**, Step length probability density function; **b**, Step length for different temporal scaling $\lambda$[1]. The steps have been normalized dividing by the average step length for every temporal rescaling. Turning points are obtained after projection along the latitude axis[2]. The pdf shows a double power-law scaling. The tail of the pdf shows a scaling domain that shrinks as the rescaling parameter increases and whose exponent depends also on the rescaling parameter decreasing from a value close to 2. For intermediate step lengths the pdf shows a scaling compatible with a power law with exponent 0.6; **c**, Angle distribution for consecutive displacements for two different time windows, 0.5 days (red) and 10 day (black).

**Figure S3.** Mean displacement (black) and root mean square displacement (red) scaling with time $T$. Both the mean displacement (black), $\langle d(T) \rangle$, and the root mean square displacement (red), $\sqrt{\langle d^2(T) \rangle}$, show the same scaling with $T$: $T^\alpha$, with $\alpha=0.83$.

**Figure S4.** Spatial density of observations. Probability density function of the number of observations per grid cell area $\rho_{ev}$.

**Figure S5.** Occupancy map of the reshuffled trajectories. Inset: comparison of the trajectories representing three realizations of the reshuffled (red, green and blue) and observed (black) locations of one individual. The colour scale is logarithmic. Figure generated with Matplotlib Basemap Toolkit[3].

**Figure S6.** Diversity of trajectories according to their normalized entropy $S$. Probability density function of the entropy of all the trajectories.

**SI Figures**

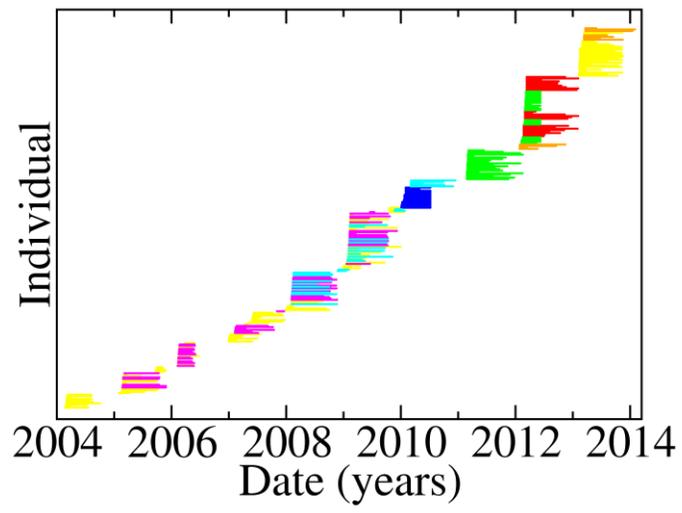

**Figure S1.**

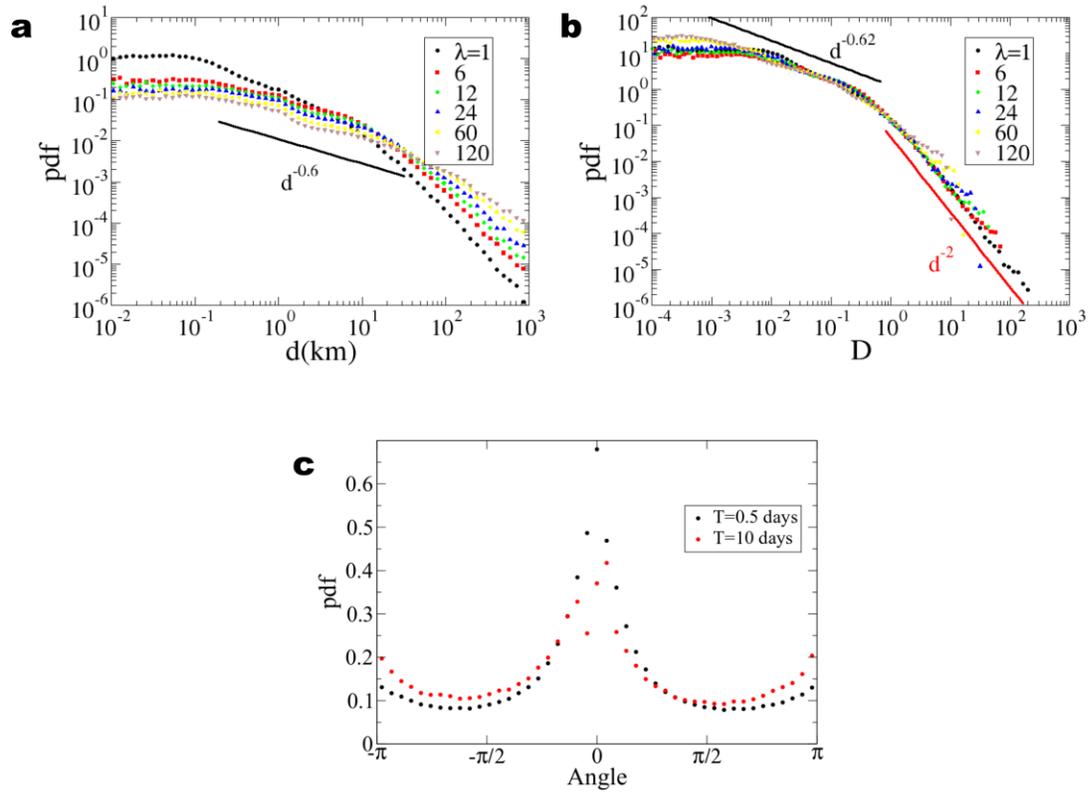

**Figure S2.**

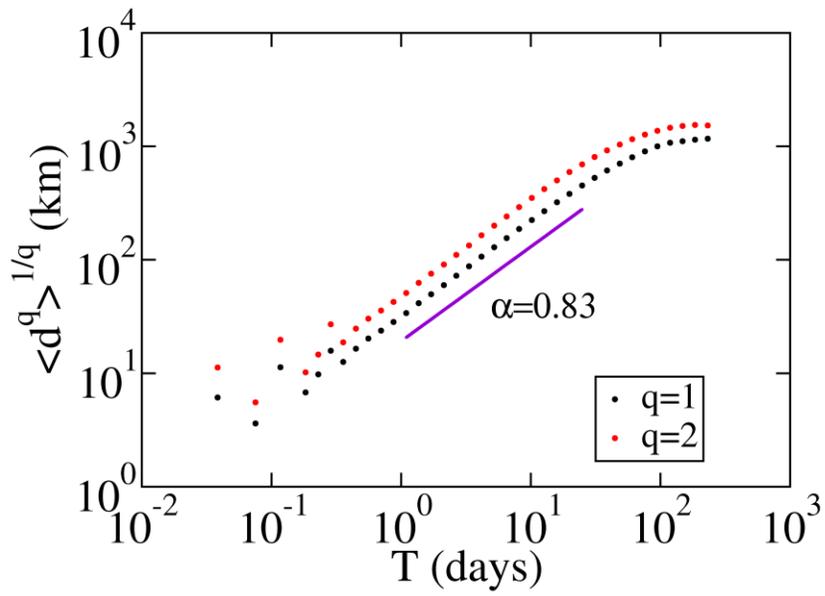

**Figure S3.**

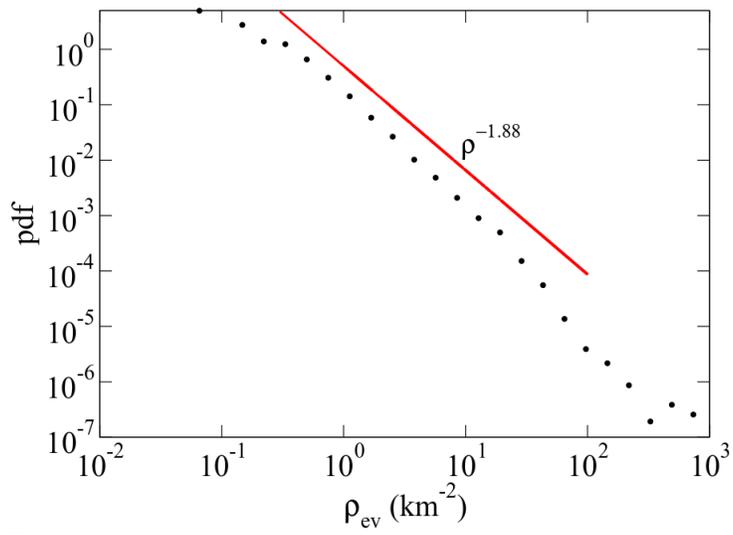

**Figure S4.**

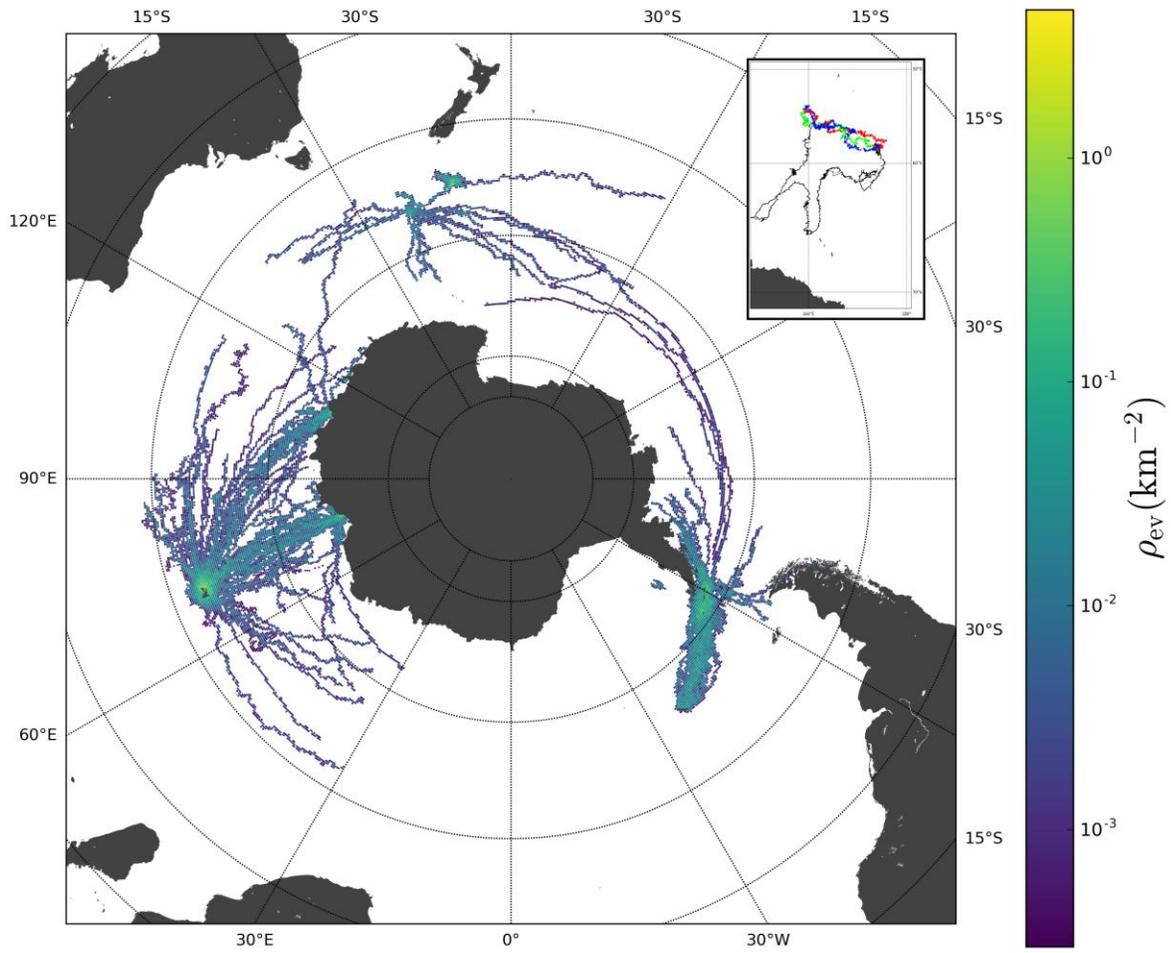

**Figure S5.**

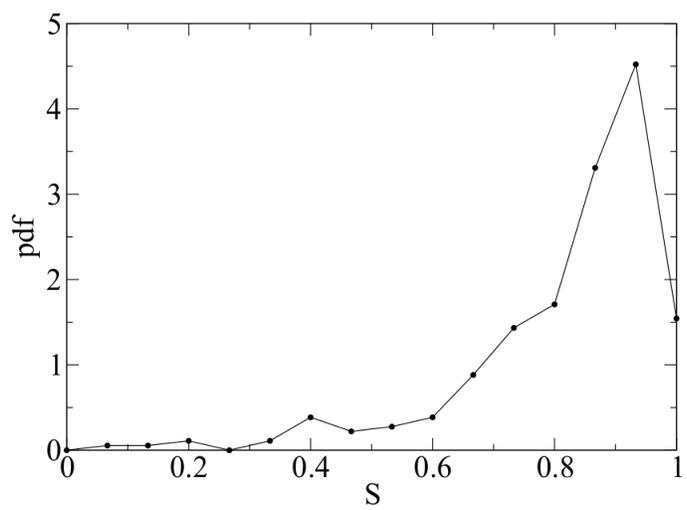

**Figure S6.**